\documentclass[aps,twocolumn,prl,bibnotes,superscriptaddress]{revtex4}
\usepackage{graphicx}

\begin{document}

\title[]{Spontaneous spin polarization in quantum point contacts}

\author{L.~P. Rokhinson}
%\email[]{leonid@physics.purdue.edu}
\affiliation{Department of Physics, Purdue
University, West Lafayette, IN 47907 USA}

\author{L.~N. Pfeiffer}
\author{K.~W. West}
\affiliation{Bell Laboratories, Lucent Technologies, Murray Hill, New Jersey
07974 USA}

\begin{abstract}
We use spatial spin separation by a magnetic focusing technique to probe the
polarization of quantum point contacts. The point contacts are fabricated from
p-type GaAs/AlGaAs heterostructures. A finite polarization is measured in the
low-density regime, when the conductance of a point contact is tuned to
$<2e^2/h$. Polarization is stronger in samples with a well defined ``0.7
structure''.
\end{abstract}

%\date{\today , ---= draft: \jobname.tex =---}
\date{Accepted to Physical Review Letters on April 4, 2006}

\maketitle

Mesoscopic systems exhibit a range of non-trivial spin-related phenomena in the
low density regime, where inter-particle Coulomb interactions become comparable
to their kinetic energy. In zero-dimensional systems the spontaneous
polarization of a few-electron quantum dot leads to spin blockade
\cite{weinman95,rokhinson01a,huttel03}, a remarkable effect where the mismatch
of a single spin blocks macroscopic current flow. In two-dimensional hole gases
there is experimental evidence of a finite spin polarization even in the
absence of a magnetic field \cite{papadakis99}. In one-dimensional systems --
quantum wires and quantum point contacts -- a puzzling so-called ``0.7
structure'' has been observed below the first quantization plateau
\cite{thomas96}. Experiments suggest
\cite{thomas96,kristensen00,cronenwett02,roche04} that an extra plateau in the
conductance vs gate voltage characteristic at $0.7\times 2e^2/h$ is spin
related, however, the origin of the phenomenon is not yet understood and is
highly debated. In this paper we report direct measurements of finite
polarization of holes in a quantum point contact (QPC) at conductances
$G<2e^2/h$. We incorporated a QPC into a magnetic focusing device
\cite{focusing} so that the polarization can be measured directly using a
recently developed spatial spin separation technique \cite{rokhinson04}.

Quantization of the ballistic conductance $G$ in integer multiples of
$g_0=2e^2/h$ is a fundamental property of 1D systems \cite{vanwees88,wharam88},
which originates from the exact cancellation of velocity and the 1D density of
states. Each energy level below the Fermi energy inside a 1D channel
contributes $0.5 g_0$ to the total conductance, and an extra factor of two
accounts for the spin degeneracy. Strong magnetic fields can lift the
degeneracy; in this case quantization in multiples of $0.5 g_0$ is observed.
This single-particle result is robust even in the presence of electron-electron
interactions because they preserve the center-of-mass velocity of the scattered
electrons. Thus, the observation of a quantized plateau at $0.7 g_0$ in the
absence of magnetic field in n-GaAs \cite{thomas96}, p-Si \cite{bargaev02},
n-GaN \cite{chou05} and p-GaAs QPCs, as well as in long clean 1D
wires\cite{dePicciotto04}, poses a serious challenge to our understanding of 1D
conductors.

Phenomenologically, the observed structure can be explained if one assumes the
existence of a {\it static} spin polarization at zero magnetic field and
confinement-dependent spin splitting of the spin subbands \cite{reilly02}.
However, the well-known Lieb--Mattis theorem forbids polarization in 1D systems
\cite{lieb62}. Some theories suggest a possible deviation from this theorem in
a realistic channel with finite width \cite{spivak00,berggren02,klironomos05}.
Recently, it has been pointed out that the temperature and bias dependence of
the differential conductance around the $0.7 g_0$ plateau are similar to the
Kondo phenomenon, thus suggesting {\it dynamic} spin
polarization\cite{cronenwett02,meir02}. Alternative theories assume {\it no
polarization} and attribute the phenomenon to electron-phonon interactions
\cite{seelig03} or to the formation of a Wigner crystal \cite{matveev04a}. None
of the above theories describe the variety of observed phenomena in a unified
and consistent fashion. Thus, direct measurement of the spin polarization
becomes of paramount importance.

\begin{figure}[t]
\def\ffile{fig1}
\centering\includegraphics[scale=1.0]{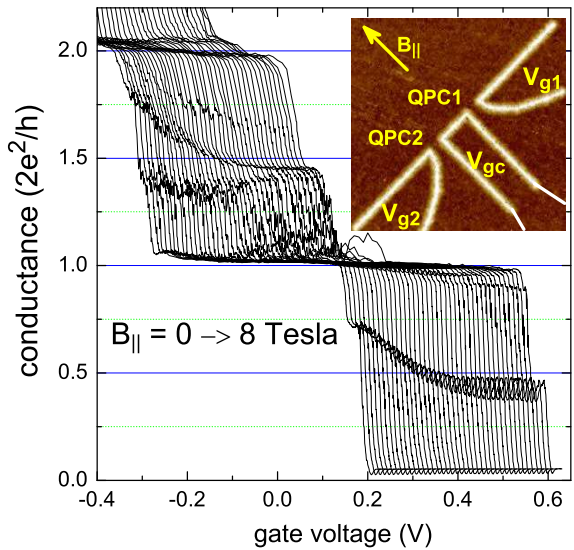} \vspace{-0.3in} \caption{ The
conductance of a quantum point contact $G$ is plotted as a function the gate
voltage $V_{g1}$  for in-plane  magnetic fields $0<B_{||}<8$ T at temperature
$T=50$ mK for QPC1. The curves are offset proportionally to $B_{||}$, the
leftmost is $B_{||}=0$. Inset: AFM micrograph of a sample ($3.3\protect\mu$m$
\times3.3\protect\mu$m). Light lines are the oxide which separates different
regions of the 2D hole gas. The two point contacts QPC1 and QPC2 form a
magnetic focusing device. The conductance of the QPCs is controlled via
voltages applied to the gates $V_{g1}$, $V_{g2}$, and the central gate
$V_{gc}$. The direction of $B_{||}$ is indicated by an arrow.}
\label{\ffile}
\end{figure}

Our devices are fabricated from a two-dimensional hole gas (2DHG) using an
atomic force microscopy local anodic oxidation technique (AFM
LAO)\cite{snow94,held98}. Oxide lines separate the 2DHG underneath by forming
$\sim200$ mV potential barriers. Several specially designed heterostructures
are grown by MBE on [113]A GaAs\cite{rokhinson02}. Despite its very close
proximity to the surface (350\AA), the 2DHG has exceptionally high mobility
$\sim 0.5\cdot10^{6}$ V$\cdot$s/cm$^2$.  The devices are fabricated from two
wafers with hole densities $p=1.47\cdot10^{11}$ cm$^{-2}$ (wafer A) and
$p=0.9\cdot10^{11}$ cm$^{-2}$ (wafer B). For quantitative analysis we use data
collected during a single cooldown for each device. The qualitative features
are reproducible upon several thermal cyclings.

The devices consist of two QPCs separated by a central gate (see inset in
Fig.~\ref{fig1}). The potential inside the point contacts can be controlled
separately by the two side gates $V_{g1}$ and $V_{g2}$, or by the central gate
$V_{gc}$. The conductance of point contact QPC1 is plotted in Fig.~\ref{fig1}
as a function of the gate voltage $V_{g1}$. At zero field (left-most curve),
plateaus with conductance quantized at $g_0$ and $2g_0$ are clearly observed.
In addition, an extra plateau can be seen at $G\sim 0.7g_0$ and, less
developed, at $G\sim 1.7g_0$. When an in-plane magnetic field $B_{||}$ is
applied, the $0.7g_0$ and $1.7g_0$ plateaus gradually shift toward $0.5g_0$ and
$1.5g_0$, saturating for $B_{||}>4$ T. This gradual decrease is different from
the abrupt appearance of half-integer plateaus for higher energy levels. In
that case the plateaus become more prominent as the Zeeman splitting increases,
but the conductance values of the plateaus do not change with $B_{||}$,
consistent with the single-particle picture.

\begin{figure}[t]
\def\ffile{fig2}
\centering\includegraphics[scale=0.95]{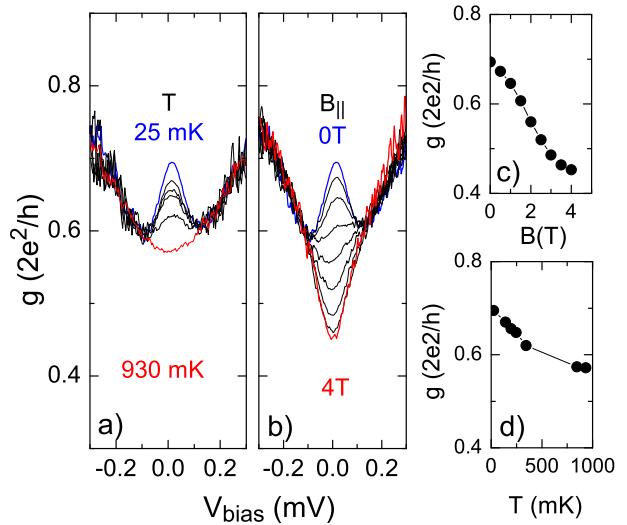} \vspace{-0.3in}
\caption{Differential conductance $g=dI/dV$ is measured
as a function of dc bias $V_{bias}$ across the QPC1. Gate voltage $V_{g1}$ is
fixed in the middle of the $0.7\times2e^2/h$ plateau at $T=25$ mK and
$B_{||}=V_{bias}=0$. In a) $B_{||}=0$ and $T=25$, 140, 190, 250, 340 and 930
mK, in b) $T=25$ mK and $B_{||}$ changes between 0 and 4 T in steps of 0.5 T.
Zero bias anomaly is the strongest at the lowest $T$ and $B_{||}=0$ and is
suppressed as $T$ and/or $B_{||}$ increases. c-d) $B$ and $T$-dependence of $g$
at $V_{bias}=0$.}
\label{\ffile}
\end{figure}

Another signature of the ``0.7 structure'' is the anomalous nonlinear
differential conductance $g=dI/dV$. A distinct peak in $g$ vs dc bias
$V_{bias}$ has been reported in electron QPCs \cite{cronenwett02}. The
nonlinear conductance in our hole device is analyzed in Fig.~\ref{fig2}.
Indeed, there is a well developed zero-bias peak at the lowest $T=25$ mK and
$B_{||}=0$. The peak is suppressed if $T$ or $B_{||}$ are increased. $g(T)$ and
$g(B_{||})$ at $V_{bias}=0$ are plotted in Fig.~\ref{fig2}(c,d). A zero-bias
peak and its suppression by $T$ and $B_{||}$ is a hallmark of the Kondo
phenomenon. The Land\'{e} factor $g^*\approx 0.3$ in the point contact is too
small to result in a detectable Zeeman splitting of the zero-bias anomaly in
our samples.

\begin{figure}[t]
\def\ffile{fig3}
\centering\includegraphics[scale=0.95]{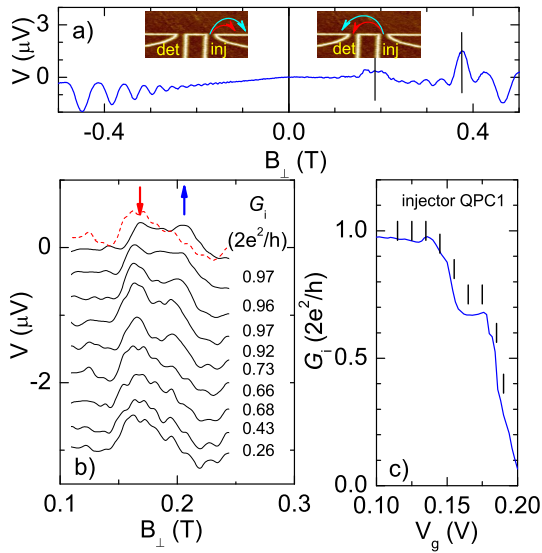} \vspace{-0.3in}
\caption{Polarization detection via magnetic focusing. a) The voltage across the
detector QPC2 is measured as a function of the perpendicular magnetic field
($B_{\bot}$). A current of 0.5 nA is flowing through the injector QPC1. The
positions of the first two magnetic focusing peaks are marked with vertical
lines. The trajectories of the ballistic holes for positive and negative
$B_{\bot}$ are shown schematically in the insets. b) The first focusing peak is
measured at different injector conductances with the detector tuned into the
middle of the $2e^2/h$ plateau. The curves are vertically offset by $-0.4$
$\mu$V relative to the top one. The $G=0.66g_0$ curve is also plotted without
an offset (red dashed line). c) THe gate voltage characteristic of QPC1.
Vertical lines mark the positions where the curves in b) are taken.}
\label{\ffile}
\end{figure}

Experimentally, it is possible to clarify the origin of the ``0.7 structure''
by measuring the polarization of carriers emerging from the QPC. The
polarization can be measured in a ballistic magnetic focusing device with
spin-orbit interaction, where carriers with opposite spin have different
cyclotron orbits in small external magnetic field \cite{rokhinson04}. In the
presence of spin-orbit interaction, carriers with the same energy and opposite
spin orientation have different momenta $\hbar
k^{\pm}=\sqrt{2m(e_f-\gamma^2/m)}\pm\gamma$, where $e_f=2\pi p\hbar^2/m$ is the
Fermi energy, $\hbar$ is Plank's constant, $p$ is 2D hole density, $m$ is the
effective mass and $\gamma$ is a spin-orbit interaction constant,
$\gamma/\hbar\approx 10^{-3}$\AA\ in our samples. Thus, in a weak orbital
magnetic field $B_{\bot}$, carriers with opposite spins injected from a QPC
will have different cyclotron radii $r_c^{\pm}=\hbar k^{\pm}/eB_{\bot}$ and can
be selectively focused into the detector QPC at $B_{\bot}^{\pm}=\hbar
k^{\pm}/2eL$ and measured separately ($L$ is the distance between the injector
and the detector QPCs).

Magneto-focusing data are plotted in Fig.~\ref{fig3}a. Current is injected
through QPC1 and the voltage is monitored across the detector QPC2. At
$|B_{\bot}|>0.25$ T, Shubnikov de-Haas oscillations in the adjacent 2D hole gas
are observed. At $B_{\bot}>0$ extra peaks due to magnetic focusing are
superimposed on Shubnikov de-Haas oscillations and the first two peaks are
clearly observed. The positions of the peaks are close to the expected values
for a QPC1-QPC2 separation of $L=0.8$ $\mu$m and scale with the 2D hole density
as $\sqrt{p_A/p_B}=1.2$ for the devices fabricated from wafers A and B. The two
peaks within the first focusing peak correspond to the focusing conditions for
the two orthogonal spin states in the 2DHG and are adiabatically related to the
pure spin states inside the point contacts \cite{rokhinson04}.

%\begin{figure}[t]
%\def\ffile{fig4}
%\centering\includegraphics[scale=1.2]{\ffile} \vspace{-0.3in} \caption{{\bf
%Magnetic focusing for different contact configurations.} Current is injected
%through either QPC1 or QPC2 and voltage is measured across the other point
%contact (blue curves). The relative strength of the peaks does not change upon
%injector-detector exchange. The QPCs are gated asymmetrically to reduce the
%width of the channel, with $V_{g2}-V_{gc}=0.46$ V (blue curve) and
%$V_{g2}-V_{gc}=0.37$ V (magenta curve) for QPC2. The degree of the asymmetry
%controls the shape of the confining potential, as shown schematically in the
%inset.}
%\label{\ffile}
%\end{figure}

\begin{figure}[t]
\def\ffile{fig5}
\centering\includegraphics[scale=1.2]{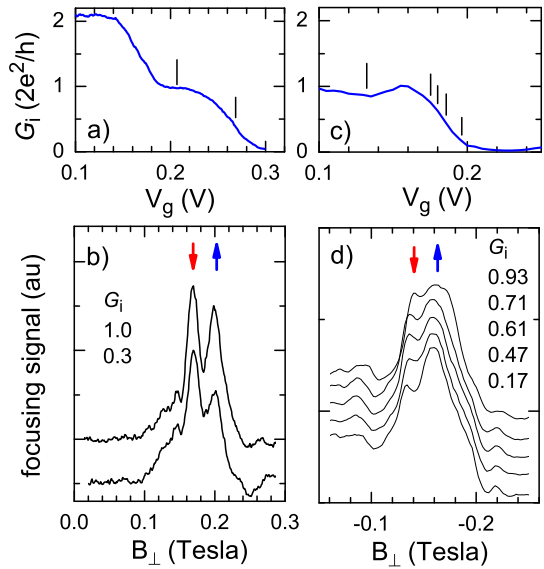} \vspace{-0.3in}
\caption{Polarization in samples with no well defined ``0.7 structure''. (a,c)
Conductance of the injector QPCs for two samples. (b,d) The first focusing peak
is plotted for fixed $G_d=2e^2/h$ and $G_i$ as indicated in the labels (in
units of $2e^2/h$). Vertical lines in (a,c) mark the positions where the
corresponding curves in (b,d) are taken. Curves in (b,d) are offset for
clarity.}
\label{\ffile}
\end{figure}

The height of the focusing peaks $V_{d}=\alpha G_{d}^{-1} I_{i} (1+P_{i}P_{d})$
is proportional to the total injector current
$I_i=I^{\uparrow}+I^{\downarrow}$, the polarization of the injector
$P_i=(I^{\uparrow}-I^{\downarrow})/(I^{\uparrow}+I^{\downarrow})$ and the
detector $P_d=(T^{\uparrow}-T^{\downarrow})/(T^{\uparrow}+T^{\downarrow})$, and
the efficiency of the current transfer $\alpha$ between the injector and the
detector \cite{potok02}. Our case of spatial spin separation corresponds to
$P_d^{\uparrow}=+1$ and $P_d^{\downarrow}=-1$ for the two peaks. Thus, the
injector polarization can be extracted as
$P_i=(V_d^{\uparrow}-V_d^{\downarrow})/(V_d^{\uparrow}+V_d^{\downarrow})$,
where $V_d^{\uparrow}$ and $V_d^{\downarrow}$ are measured at the same $G_i$,
or, alternatively, as $P_i=|V_d^{\sigma}/V_{d0}^{\sigma}-1|$ for each peak,
where $V_{d0}^{\sigma}$ is the corresponding peak height for unpolarized
injection.

The dependence of the first focusing peak on the injector conductance is shown
in Fig.~\ref{fig3}b. The top curve is measured with the conductance of both
QPC1 and QPC2 tuned into the first quantized plateau $G=g_0=2e^2/h$. Both peaks
have approximately the same value, consistent with the expectation that at
$G=g_0$ there are two fully transmitting spin states below the Fermi energy. We
fix the detector QPC2 at $G_d=g_0$ to allow both spin states to be detected and
gradually reduce the conductance of the injector QPC1 to $G_i<g_0$. As $G$
decreases, the height of the high-$B$ peak within the first focusing peak
decreases, while the height of the low-$B$ peak increases. This indicates that
the two subbands with opposite spins are not equally populated at $G<g_0$ and,
thus, there is a finite polarization of holes injected from QPC1. We estimate
$P_i\approx40\pm15$\% for $G<0.9g_0$ using either the ratio of the two peaks or
suppression/enhancement of each peak. Note that the polarization due to Zeeman
splitting of the spin subbands in an external magnetic field is too small to be
detected in our experiments, $g^*\mu_BB_{\bot}\approx 6\ \mu\text{eV}\lesssim
k_B T, eV_{ac}$. Also, we do not expect the hyper-fine interaction to play a
significant role since the leading contact--Fermi term is absent for holes.

The appearance of a plateau around $0.7 g_0$ requires substantial energy
splitting between the two spin subbands, comparable to or larger than the level
broadening. In many QPCs, though, this condition is not met and there is no
extra plateau below $g_0$. The question remains whether there is still a finite
polarization below the first quantized plateau. We investigated several QPCs
with no ``0.7 structure'', see Fig.~\ref{fig5}. The samples are fabricated from
different wafers A (left panel) and B (right panel). The injector QPCs in both
devices have well defined first quantized plateau at $2e^2/h$ but no ``0.7
structure''. The magnetic focusing signal is measured with the detector QPC
fixed at $G_d=g_0$. At $G_{i}=g_0$ the first focusing peak is split in two
peaks of similar height, with both spin subbands being populated. As $G_{i}$ is
decreased below $g_0$, one of the peaks becomes suppressed while the other is
enhanced, similar to the device with a well defined ``0.7 structure''. The
polarization $P_i$ increases gradually from 0 to $\sim15$\% as $G_i$ decreases
from $1g_0$ to $0.2g_0$; the polarization is approximately two times lower than
in the device with the ``0.7 structure''. We conclude that the polarization of
QPCs near the onset of conduction is a rather generic property and that the
appearance of the ``0.7 structure'' is an extreme indicator of such
polarization when the spin gap becomes large enough to result in a measurable
feature in the gate voltage characteristic.

The two devices in Fig.~\ref{fig5} have different crystallographic orientations
and, thus, different angles between the momentum of the injected carriers and
the internal spin-orbit field ($I_i || [\overline{2}33]$ and $I_i ||
[0\overline{1}1]$ for the samples on the left and right panels, respectively)
which, presumably, results in a different peak being suppressed. For the device
to work as a spin detector it is sufficient that each spin state in a QPC
adiabatically maps onto one of the chiral states in the adjacent 2D gas, which
has been checked by the application of a strong Zeeman field as discussed in
Ref.~\cite{rokhinson04}. The exact mapping conditions are the subject of
ongoing research.

We conducted several tests to ensure that the extracted polarization is not
dominated by disorder-mediated fluctuations. The reported data were
reproducible over several thermal cyclings to room temperature (six for the
sample in Fig.~\ref{fig3}). Switching the injector and detector with a
simultaneous reversal of the magnetic field results in almost identical
magnetic focusing data. The $P_i$ calculated from each peak at the same $B$ is
consistent with the $P_i$ calculated from both peaks $\Delta B=30$ mT apart,
which is comparable with the period of the mesoscopic fluctuations in similar
structures. Asymmetric gating of the point contact shifts the conducting
channel in space and, thus, allows us to scan through the underlying disorder
potential \cite{heinzel00}. Changing $V_{gc}-V_{g1}$ by 90 mV shifts the
channel by $\approx 7$ nm, while the correlation length for the disorder inside
a 1D channel in similar but higher mobility electron samples was measured
$\approx 2$ nm. In our sample this shift also translates into an extra
half-flux quantum being inserted inside the focusing trajectory.
Experimentally, the peak heights remain the same as we laterally shift the
injector channel (although the peaks become slightly broader). Finally, the
peak height is sensitive to in-plane magnetic field (see Fig. 2 of
Ref.~\onlinecite{rokhinson04}), which is expected for spin subbands but not for
mesoscopic fluctuations. Thus, our experiments provide a direct measurement of
finite polarization in point contacts.

In conclusion, we present an experimental investigation of the ``0.7
structure'' in p-type QPCs with a new twist: a direct measurement of the spin
polarization. Using a newly developed spin separation technique, we determine
the polarization of the holes injected from a QPC into an adjacent 2D gas. The
technique is sensitive to {\it static} polarization, which is found to be as
high as 40\% in samples with a well defined `0.7 structure''. Some polarization
has been measured in all point contacts below the first plateau. This result
questions the Kondo interpretation as an origin of the ``0.7 structure'', which
is incompatible with a finite static polarization. The ``0.7 structure'' in
p-type QPCs shows all the essential features reported for n-type QPCs, such as
a gradual evolution into the $0.5 g_0$ plateau at high in-plane magnetic
fields, survival at high temperatures, a gradual increase toward $1.0 g_0$ at
low temperatures, and the zero-bias anomaly, which is suppressed by either
temperature increase or application of a magnetic filed.  The similarities
between p-type and n-type QPCs suggest that the underlying physics responsible
for the appearance of the ``0.7 structure'' should be the same.

Authors thank Yu.~Lyanda-Geller and G.~F. Giuliani for discussions. This work
was supported by NSF grant ECS-0348289.

\bibliographystyle{revtex}
\bibliography{rohi,07anomaly}

\end{document}